\documentclass[twocolumn,showpacs,preprintnumbers,amsmath,amssymb]{revtex4}

\usepackage{graphicx}
\usepackage{dcolumn}
\usepackage{bm}

\begin{document}


\title{Resonant Two-Magnon Raman Scattering and Photoexcited States in Two-Dimensional Mott Insulators}

\author{T. Tohyama}
\author{H. Onodera}
\author{K. Tsutsui}
\author{S. Maekawa}
\affiliation{Institute for Materials Research, Tohoku University, Sendai 980-8577, Japan}

\date{\today}

\begin{abstract}
We investigate the resonant two-magnon Raman scattering in two-dimensional (2D) Mott insulators by using a half-filled 2D Hubbard model in the strong coupling limit.  By performing numerical diagonalization calculations for small clusters, we find that the Raman intensity is enhanced when the incoming photon energy is not near the optical absorption edge but well above it, being consistent with experimental data.  The absence of resonance near the gap edge is associated with the presence of background spins, while photoexcited states for resonance are found to be characterized by the charge degree of freedom.  The resonance mechanism is different from those proposed previously.
\end{abstract}

\maketitle

Two-dimensional (2D) insulating cuprates are typical Mott insulators.   The understanding of their electronic states is of crucial importance since they are parent compounds of high-$T_c$ cuprate superconductors.  In the Mott insulators, low-energy physics associated with the spin degree of freedom is described by a 2D Heisenberg model.  However, the nature of high-energy excited states across the Mott gap is less clear as compared with that of the low-energy spin states.

Such excited states are usually investigated by optical absorption measurements, where photons excite the system across the Mott gap.  In addition to the optical absorption, resonant Raman scattering is useful for the investigation of the photoexcited states.  In the scattering, resonant effects are observed as enhancements of Raman scattering intensities associated with phonons and magnons when the incoming photon energy $\omega_{\rm i}$ is tuned near the Mott gap.  In the 2D insulating cuprates such as La$_2$CuO$_4$, YBa$_2$Cu$_3$O$_6$, and Sr$_2$CuO$_2$Cl$_2$, the $\omega_{\rm i}$ dependence of two-magnon (2M) scattering intensity with $B_1$ scattering geometry shows an interesting feature~\cite{Yoshida,Blumberg}: there is no resonance at a peak near the gap edge in the absorption spectra ($\sim$1.8$-$2~eV) but the resonance occurs at around $\omega_{\rm i}\sim$3~eV, well above the absorption-edge peak.  Since such $\omega_{\rm i}$ dependence of the intensity is not a common behavior of transition metal oxides~\cite{Merlin}, novel properties of the photoexcited states are expected in the 2D Mott insulators.
In contrast to a large number of theoretical studies about the asymmetric line shape of the $B_1$ 2M Raman spectrum~\cite{Sandvik,Freitas}, there are a smaller number of theoretical works on the $\omega_{\rm i}$ dependence of the intensity~\cite{Shastry,Chubukov,Schonfeld,Hanamura}.  As the origin of the resonance, a triple resonance mechanism based on a spin-density-wave formalism~\cite{Chubukov,Schonfeld} and an excitonic mechanism in which the charge-transfer (CT) exciton is taken into account~\cite{Hanamura} have been proposed so far.

In this Letter, we propose a new mechanism of the resonant 2M Raman scattering and clarify the nature of the photoexcited states in 2D Mott insulators.  Numerical calculations of the 2M Raman intensity for various scattering geometries as a function of $\omega_{\rm i}$ are performed for a half-filled 2D Hubbard model in the strong coupling limit.  We find that the model can reproduce the experimental observations where the resonance in the $B_1$ geometry occurs when $\omega_{\rm i}$ is not near the Mott gap edge but well above it.  The absence of resonance near the gap edge is found to be closely associated with the presence of background spins.  On the other hand, the photoexcited states where the resonance occurs show less importance of the spin degree of freedom.  Rather, the spatial distribution of charge carriers created by photons plays a dominant role in the resonance.  It is crucial to notice that these conclusions are obtained by treating the Hubbard model exactly, taking into account full configurations associated with both the charge and spin degrees of freedom, in contrast to the previous works~\cite{Chubukov,Schonfeld,Hanamura}.  Two-magnon Raman intensities with $A_1$, $B_2$, and $A_2$ geometries are found to show resonance behaviors that are different from those of $B_1$.

Insulating cuprates are known to be CT-type Mott insulators, where both Cu3$d$ and O2$p$ orbitals participate in the electronic states.  However, it is well established that the electronic states of the CT-type insulators can be described by a Hubbard model with a half-filled single band~\cite{Maekawa}.  The Hubbard Hamiltonian with the nearest neighbor (NN) hopping in 2D is given by
$
H_{\rm Hub}=-t\sum_{\langle i,j\rangle, \sigma} \left( c_{i,\sigma}^\dagger c_{j,\sigma}+ {\rm H.c.} \right)
+U\sum_i n_{i,\uparrow}n_{i,\downarrow}
$,
where $c_{i,\sigma}^\dagger$ is the creation operator of an electron with spin $\sigma$ at site $i$, $n_{i,\sigma}=c_{i,\sigma}^{\dagger}c_{i,\sigma}$, $\langle i,j\rangle$ runs over pairs on the NN sites, $t$ is the NN hopping integral, and $U$ is the on-site Coulomb interaction.  The value of $t$ is estimated to be $t\sim0.35$ eV~\cite{Maekawa}.  The value of $U$ is estimated to be $U=$10$t$ for the gap values to be consistent with experimental ones.

In the strong coupling limit ($U \gg t$), the ground state at half filling has one spin per site; i.e., there is no doubly occupied site.  In this case, the low-energy excitation of the Hubbard model may be described by a Heisenberg model: $H_0=J\sum_{\langle i,j \rangle} \left( {\bm S}_i \cdot {\bm S}_j - \frac{1}{4} n_i n_j \right)$, where ${\bm S}_i$ is the spin operator with $S=1/2$ at site $i$, $n_i$=$n_{i,\uparrow}$+$n_{i,\downarrow}$, and $J=4t^2/U$.  On the other hand, the photoexcited states have both one doubly occupied site and one vacant site.  In order to obtain an effective Hamiltonian describing the photoexcited states, we restrict the Hilbert spaces to a subspace with one doubly occupied site.  Performing the second order perturbation with respect to the hopping term $H_t$ in the Hubbard Hamiltonian, the effective Hamiltonian is given by
\begin{equation}
H_{\rm eff}=\Pi_1H_t\Pi_1-\frac{1}{U}\Pi_1H_t\Pi_2H_t\Pi_1+\frac{1}{U}\Pi_1H_t\Pi_0H_t\Pi_1+U,
\label{Heff}
\end{equation}
where $\Pi_0$, $\Pi_1$, and $\Pi_2$ are projection operators onto the Hilbert space with zero, one, and two doubly occupied sites, respectively.  A complete expression of Eq.~(\ref{Heff}) has been given elsewhere~\cite{Takahashi}.

At zero temperature, the Raman scattering intensity is expressed in a Fermi's golden rule form,
$
R(\omega)=\sum_f \left| \langle f \left| M_R \right| i \rangle \right|^2 \delta \left( \omega -E_f + E_i \right)
$,
where $\left| f \right\rangle$ denotes a final state with energy $E_f$ and $\left| i \right\rangle$ is an initial state with energy $E_i$.  In the present case, $\left| i \right\rangle$ is just the ground state of the Heisenberg Hamiltonian $H_0$, while $\left| f \right\rangle$ is an excited state of $H_0$ with two magnons.  The Raman shift is given by $\omega=\omega_{\rm i} - \omega_{\rm f}$, $\omega_{\rm f}$ being the outgoing photon energy.  $M_R$ is the Raman tensor operator.  We first consider the $B_1$ representation of the $C_{4v}$ group because its contribution dominates 2M Raman intensity in 2D Mott insulators~\cite{Shastry,Sulewski}.  The dominant contribution to the $B_1$ Raman spectrum comes from a matrix element,
\begin{equation}
\langle f \left| M_R^{B_1} \right| i \rangle = \frac{1}{2} \sum_\mu \left[ \frac{ \langle f \left| j_x \right| \mu \rangle \langle \mu \left| j_x \right| i \rangle - \langle f \left| j_y \right| \mu \rangle \langle \mu \left| j_y \right| i \rangle}{E_\mu - E_i -\omega_{\rm i} + i\Gamma} \right]
\label{B1g}
\end{equation}
with the current operator $j_\alpha = i t \sum_{i,\sigma} ( \tilde{c}^\dagger_{i+\alpha,\sigma} \tilde{c}^{}_{i,\sigma} - \tilde{c}^\dagger_{i,\sigma} \tilde{c}^{}_{i+\alpha,\sigma} )$.  Here, the intermediate state $\left| \mu \right\rangle$ is an eigenstate of $H_{\rm eff}$ with the eigenenergy $E_\mu$ and $\Gamma$ is a damping rate for $\left| \mu \right\rangle$.  We note that the creation and annihilation operators, $\tilde{c}^\dagger_{i,\sigma}$ and $\tilde{c}^{}_{i,\sigma}$, in the current operator are projected onto the subspaces with either zero or one doubly occupied site.

We perform numerical calculations of the $B_1$ Raman spectrum for a $\sqrt{20}\times\sqrt{20}$ square cluster with periodic boundary conditions.  The Heisenberg ground state is calculated by the Lancz\"os method, and the spectrum $R(\omega)$ is obtained by utilizing a modified version of the conjugate-gradient method together with the Lancz\"os technique.

\begin{figure}
\begin{center}
\includegraphics[width=7.6cm]{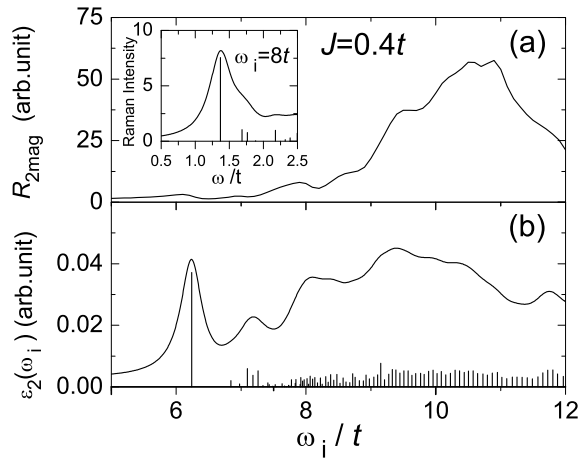}
\caption{\label{fig1}
The dependence of (a) two-magnon Raman intensity $R_{\rm 2mag}$ with $B_1$ geometry and (b) absorption spectrum $\epsilon_2$ on incoming photon energy $\omega_{\rm i}$ in a 20-site cluster with $U/t$=10 ($J/t$=0.4).  Inset of (a) shows two-magnon Raman spectrum at $\omega_{\rm i}$=8$t$.  $R_{\rm 2mag}$ represents the weight of the lowest-energy peak at $\omega$=1.37$t$ in the inset.  The solid line in (b) is obtained by performing a Lorentzian broadening with a width of 0.4$t$ on the delta functions denoted by vertical bars.
}
\end{center}
\end{figure}

The inset of Fig.~1(a) shows a 2M Raman scattering spectrum with $U/t$=10 ($J/t$=0.4), where $\omega_{\rm i}$=8$t$ is chosen to be consistent with experimental conditions~\cite{Sulewski}.  The spectrum is dominated by a large peak at $\omega$=1.37$t$=3.43$J$ accompanied by higher-energy small ones at $\omega\sim$1.7$t$, being similar to that obtained by using a standard Raman spin operator~\cite{Sandvik}.  The spectral weight changes with $\omega_{\rm i}$ without notable change of line shape.  The spectral weight of the $\omega$=1.37$t$ peak, denoted as $R_{\rm 2mag}$, is plotted as a function of $\omega_{\rm i}$ in Fig.~1(a).  $R_{\rm 2mag}$ shows maximum resonance at $\omega_{\rm i}$$\sim $11$t$.  Comparing the $\omega_{\rm i}$ dependence with the optical absorption spectrum $\epsilon_2(\omega_{\rm i})$, given by $\omega_{\rm i}^{-2}\sum_\mu \left| \langle \mu \left| j_x \right| i \rangle \right|^2 \delta(\omega_{\rm i} -E_\mu +E_i)$, we find that $R_{\rm 2mag}$ is not enhanced when $\omega_{\rm i}$ is tuned to the absorption-edge peak.  Rather, the resonance occurs when $\omega_{\rm i}$ is inside a continuum well above the Mott gap.  The $\omega_{\rm i}$ dependence of $R_{\rm 2mag}$ is thus similar to the experimental data mentioned above~\cite{Yoshida,Blumberg}.  The energy difference between the absorption-edge and the resonance maximum is about 4$t$$\sim$1.4~eV, being also consistent with experimental data with the difference of $\sim$1.35~eV~\cite{Blumberg}.  We note that, in addition to the 20-site cluster, smaller clusters with 4$\times$4 and $\sqrt{18}$$\times$$\sqrt{18}$ sites show similar dependences (not shown here).  We also checked that inclusion of longer-range hopping and Coulomb terms, which are used for a precise description of cuprates, does not change the above results~\cite{Onodera1,Footnote}.

The 2D Mott insulators before photoexcitation have localized spins interacting with each other via the antiferromagnetic (AF) exchange interaction $J$.  Charge carriers introduced into the insulators are known to induce a spin cloud around the carriers as a consequence of the misaligned spins along the carrier-hopping paths.  It is thus natural to expect that the optical responses are strongly influenced by the presence of the localized spins.  This can be checked by changing the value of the exchange interaction $J$/$t$.  In Fig.~2, the absorption spectrum multiplied by $\omega^2_{\rm i}$ (solid lines), $\alpha(\omega_{\rm i})=\omega_{\rm i}^2\epsilon_2(\omega_{\rm i})$, is shown for various values of $J$, together with the 2M Raman intensity $R_{\rm 2mag}$ (dotted lines).  In the figure, the energy is shifted by $U=4t^2/J$.  The absorption-edge peak is found to be very sensitive to the value of $J$: With decreasing $J$, the peak intensity decreases, while the resonant enhancement of $R_{\rm 2mag}$ remains even for small $J$, keeping the energy difference between the resonance position and the absorption edge almost constant.  We note that such $J$ dependence of the resonance is different from that expected from a triple resonance mechanism of the 2M Raman scattering proposed based on a spin-density-wave (SDW) formalism for the half-filled 2D Hubbard model~\cite{Chubukov,Schonfeld}, where a maximum of resonance occurs near the upper edge of the absorption band with a width of 8$J$ and thus the resonance energy must approach the absorption-edge with decreasing $J$.  This implies that the origin of resonance we have found is different from the triple resonance seen in the SDW approach.

\begin{figure}
\begin{center}
\includegraphics[width=7.9cm]{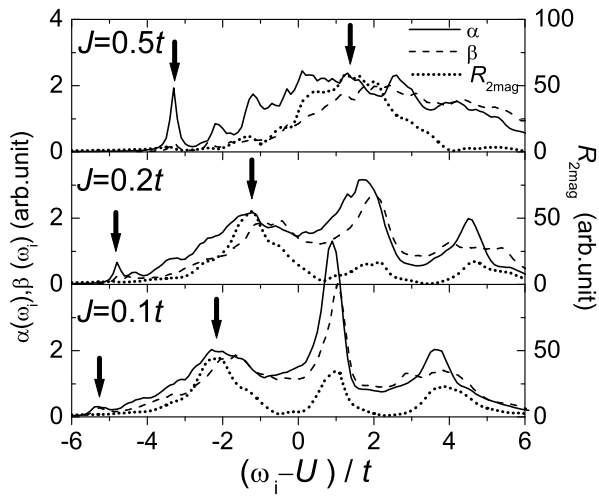}
\caption{\label{fig2}
Absorption spectra $\alpha(\omega_{\rm i})$ and $\beta(\omega_{\rm i})$ from intital (solid lines) and final (dashed lines) states of the Raman process, respectively, together with two-magnon Raman intensity $R_{\rm 2mag}$ (dotted lines) in a 20-site cluster for three $J/t$ values.  The incoming photon energy is shifted by corresponding value of $U$.  The arrows denote positions of the absorption-edge peak and the maximum of $R_{\rm 2mag}$.}
\end{center}
\end{figure}

The sensitivity of the absorption-edge peak to $J$ seen in Fig.~2 is easily understood in the following way.  The initial state, i.e., the ground state of the Heisenberg model, is dominated by N\'eel-type spin configurations inducing AF order.  The electric filed applied to the initial state creates a pair of empty and doubly occupied sites with one-lattice spacing, keeping the N\'eel-type spin configurations dominant in the spin background.  The intensity of the absorption spectrum, which is given by $\left| \langle \mu \left| j_x \right| i \rangle \right|^2$, is thus predominantly governed by two factors in the photoexcited states $\left| \mu \right\rangle$:  (i) the probability of finding the two photocreated sites with one-lattice spacing, and (ii) the probability of the N\'eel-type configurations.  In the photoexcited state at the absorption-edge peak $\left| \mu_{\rm edge} \right\rangle$, the probability (i) is not largest among other possible two-site configurations with longer distances.  Therefore, the N\'eel-type configurations in the spin background are crucial for the peak intensity.  In fact, evaluating the wave function of $\left| \mu_{\rm edge} \right\rangle$, we found that the probability (ii) increases with increasing $J$ as expected.  Therefore, at large $J$, the N\'eel configurations contribute to enhancing the matrix element $\langle \mu_{\rm edge} \left| j_x \right| i \rangle$, leading to the large absorption-edge peak.

Using this consideration, it is easy to understand why the absorption-edge peak does not resonate with the 2M Raman scattering.  In the Raman process, the matrix element of $\langle f \left| j_x \right| \mu \rangle$ also contributes to the intensity, where $\left| f \right\rangle$ is the 2M final state.  The spin correlation in $\left| f \right\rangle$ is different from that in the initial state because two magnons are excited.  By examining the wave function of $\left| f \right\rangle$, we found that the weight of the N\'eel-type spin configurations are very small in $\left| f \right\rangle$.  Therefore, the matrix element $\langle f \left| j_x \right| \mu_{\rm edge} \rangle$ is very small, resulting in no resonance at the absorption edge.  In order to confirm this, we plot in Fig.~2 an absorption spectrum from $\left| f \right\rangle$ (dashed lines), defined as $\beta(\omega_{\rm i})=\sum_\mu \left| \langle \mu \left| j_x \right| f \rangle \right|^2 \delta(\omega_{\rm i} -E_\mu +E_i)$.  As expected, it exhibits no enhancement at the photoexcited state where $\alpha(\omega_{\rm i})$ shows the absorption-edge peak at $(\omega_{\rm i}-U)/t$=$-$3.4 for $J$/$t$=0.5.

In this context, recent resonant 2M Raman scattering experiments performed on a ladder-type cuprate are interesting because the $\omega_{\rm i}$ dependence of the Raman intensity shows the same behavior as the absorption spectrum~\cite{Gozar} in contrast to the 2D systems.  In the ladder system, the difference between the initial and final states is expected to be smaller than that in 2D because of a spin liquid ground state in the ladder.  This may explain the difference between 2D and ladder~\cite{Onodera2}.

In contrast to the absorption-edge peak, a broad-peak structure, where the 2M Raman scattering resonates, is insensitive to $J$ as shown in Fig.~2.  This implies that the spin background has less effect on the structure and thus the relative position of the empty and doubly occupied sites, i.e., the probability (ii) mentioned above, is important.  In fact, by examining the two-site correlation with one-lattice spacing in all of the photoexcited states for a 4$\times$4 cluster, we found that the correlation shows an enhancement around the broad peak structure (not shown here).  The fact that $\beta(\omega_{\rm i})$ in Fig.~2 exhibits a similar broad-peak structure at the same energy region, indicated by arrows, also implies little effect of the spin degree of freedom on the corresponding photoexcited states and thus supports the idea that the photoexcited states for the 2M Raman resonance are characterized by the charge degree of freedom.

\begin{figure}
\begin{center}
\includegraphics[width=7.9cm]{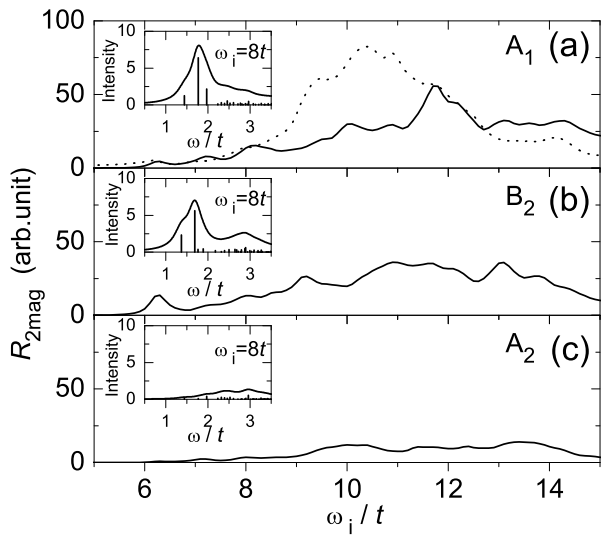}
\caption{\label{fig3}
The dependence of integrated two-magnon Raman intensities up to $\omega$=3.5$t$ for (a) $A_1$, (b) $B_2$, and (c) $A_2$ geometries on incoming photon energy $\omega_{\rm i}$ in a 20-site cluster with $U/t$=10 ($J/t$=0.4).  Insets show two-magnon Raman spectrum at $\omega_{\rm i}$=8$t$, where solid lines are obtained by performing a Lorentzian broadening with a width of 0.4$t$ on the delta functions denoted by vertical bars.  The dotted line in (a) represents integrated intensity of $B_1$ up to $\omega$=3.5$t$.
}
\end{center}
\end{figure}

Finally, in order to clarify how the resonant behaviors mentioned above are influenced by scattering geometry, we examine 2M Raman spectrum for $A_1$, $B_2$, and $A_2$ symmetries.  The corresponding Raman operators contain the current operators with the form of $j_xj_x+j_yj_y$, $j_xj_y+j_yj_x$, and $j_xj_y-j_yj_x$ for the $A_1$, $B_2$, and $A_2$, respectively.  The insets of Fig.~3 show the Raman spectra at $\omega_{\rm i}$=8$t$.  The intensity maximum in both the $A_1$ and $B_2$ Raman spectra appears at $\omega$$\sim$1.7$t$, which is higher than that of $B_1$ (1.37$t$) in Fig.~1.  In addition, the spectra are broader than that of $B_1$.  These features are qualitatively similar to experimental data~\cite{Sulewski}.  The $A_2$ Raman spectrum that reflects chiral spin fluctuations~\cite{Shastry} shows a very broad maximum at higher-energy $\omega$=3$t$, being also consistent with the experiments~\cite{Sulewski}.  In the main panels, the $\omega_{\rm i}$ dependence of the integrated Raman spectra up to $\omega$=3.5$t$ is shown.  For comparison, the integrated $B_1$ intensity up to $\omega$=3.5$t$ is plotted in Fig.~3(a) as a dotted line.  It is found that there is no pronounced resonance at the absorption edge of $\omega_{\rm i}$=6.2$t$, being similar to the $B_1$ geometry.  On the other hand, the resonance behaviors at higher $\omega_{\rm i}$ are different from that of $B_1$: the resonance maxima do not appear at $\omega_{\rm i}$$\sim$10$t$ and the intensities are smaller than that of $B_1$ geometry.  Systematic experimental studies of resonant 2M Raman scattering decomposed into various scattering geometries for a wide range of $\omega_{\rm i}$ are necessary in order to confirm these differences.  We note that, from detailed analyses that are similar to those done for the $B_1$ geometry, the main origin for the resonance maxima is found to be the same as that of $B_1$, i.e., the charge degree of freedom.  For example, a resonance maximum at $\omega_{\rm i}$$\sim$12$t$ in $A_1$ remains even for small values of $J$ and the energy of the maximum approaches the center of the photoexcited band ($\omega_{\rm i}=U$), where the probability of finding the two photocreated sites with one-lattice spacing is very large.

In summary, we have clarified the mechanism of the resonant two-magnon Raman scattering and the nature of the photoexcited states in 2D Mott insulators.  Numerical calculations for a half-filled 2D Hubbard model in the strong coupling limit have clearly reproduced the experimental observations that the resonance occurs when the incoming photon energy is well above the Mott gap energy, but does not when the energy is at the absorption-edge peak.  The spin degree of freedom is found to play a crucial role in the absence of resonance at the peak.  In particular, the difference of spin configurations between the initial and final states is a main reason for the absence resonance.  In contrast, the photoexcited states where the resonance occurs show less influence of the spin degree of freedom.  Rather, the spatial distribution of charge carriers created by photons plays a dominant role in the resonance.  In other words, the charge degree of freedom controls the resonance of the two-magnon Raman scattering in the 2D Mott insulators.  The mechanism of the resonance and nature of the photoexcited states that we have found are different from those proposed previously~\cite{Chubukov,Schonfeld,Hanamura}.  It should be emphasized that, in contrast with the previous works, our conclusions are obtained by treating the Hubbard model exactly, taking into account full configurations associated with both the charge and spin degrees of freedom.

This work was supported by a Grant-in-Aid for scientific Research from the Ministry of Education, Culture, Sports, Science and Technology of Japan, and CREST.  The numerical calculations were performed in the supercomputing facilities in ISSP, University of Tokyo, and IMR, Tohoku University.


\begin{thebibliography}{99}
\bibitem{Footnote} The consistency between the effective model and the original Hubbard model has also been confirmed for a 4$\times$4 cluster.
\end{thebibliography}
\end{document}